\documentstyle[12pt]{article}
\newcommand{\tfrac}[2]{{\textstyle{\frac{#1}{#2}}}}
\renewcommand{\title}[1]{\null\vspace{15mm}

\noindent{\Large{\bf #1}} \vspace{18mm}

}
\newcommand{\authors}[1]{\noindent{\large By} \hspace{3mm}
{\large #1}\vspace{25mm}

}
\newcommand{\address}[1]{\noindent #1\vspace{5mm}

}
\renewcommand{\abstract}[1]{\vspace{30mm}

\noindent{\small{\em Abstract.} #1}\vspace{2mm}
}
\begin{document}

\setcounter{page}{0}
\thispagestyle{empty}\hspace*{\fill}REF. TUW 95-08
\title{Superdiffeomorphisms of the Topological
\\Yang-Mills and Renormalization}
%\begin{center}
\authors{H. Zerrouki\footnote{Work supported in part by the
``Fonds zur F\"orderung der Wissenschaftlichen Forschung'',
under Contract Grant Number P10268 - PHY.}}%\\
%\end{center}

%\begin{center}
\address{Institut f\"ur Theoretische Physik,\\
         Technische Universit\"at Wien\\
         Wiedner Hauptstra\ss e 8-10,\\
         A-1040, Vienna, AUSTRIA} %\\
%\end{center}
\abstract{We give the superdiffeomorphisms transformations of the
four-dimensional topological Yang-Mills theory in curved manifold
and we discuss the ultraviolet renormalization of the model.
The explicit expression of the most general counterterm
is given.}

\newpage
\section{Introduction}

Some years ago Witten introduced the topological
Yang-Mills model \cite{wit} in order to give a quantum field
theoretical interpretation of the Donaldson polynomials \cite{dona}.
These polynomials are global topological invariants, and
correspond to the observables of a general covariant field
theory on a four manifold, which can be viewed as a twisted
version of $N = 2$ supersymmetric Yang-Mills field theory. \\
A further step in this direction was accomplished by the
authors of \cite{bs}, \cite{LPer} and \cite{osb}. Indeed, they
reformulated the model in the BRS \cite{brs} framework,
which, after gauge fixing, yields to the topological Yang-Mills
theory. And in fact it was observed that such theory arise
after gauge-fixing a topological invariant term.
The renormalization of such a theory was considered in
\cite{bms}
%, \cite{h}, \cite{brt}, \cite{bbt}, \cite{M}, \cite{mwo}
%and
-- \cite{why}.

The physical motivation behind studying the topological
quantum field theories (see \cite{tft} for a general review),
is that they may give us a description of a phase of
unbrocken diffeomorphisms invariance in quantum gravity.

One motivation of the present paper is to analyse the
ultraviolet behaviour of the topological Yang-Mill theory
on a curved four dimensional manifold by making use of the
superdiffeomorphisms symmetry.
This is done in two steps: first we extend the classical
analysis of \cite{why} to a four dimensional Riemannian manifold,
constructing the superdiffeomorphism transformations, which are
nothing else than a generalisation of the vector supersymmetry
in a curved space-time\cite{lp}, \cite{mwosz}.
And this will make the subject of section $2$.

Second, we discuss the ultraviolet behaviour of the
topological Yang-Mills, where
we will use the algebraic renormalization techniques, which suppose
the existence of a substraction scheme. This is not true for any curved
manifold, but at least such substraction scheme exists
in manifolds which are topologicaly equivalent to the four dimensional
flat space-time and endowed with asymtotically flat metric.
As the topology does not modify the
short-distance (the ultraviolet) behaviour of the theory, then in our
analysis we expect the same ultraviolet properties as in the flat
space--time limit \cite{why}. This will be done in section $3$.
The paper ends with a conclusion.

\newpage

\section{The Classical analysis in the Landau Gauge}

We devote this section to the investigation of the properties of the
topological Yang-Mills theory on a curved Riemannian manifold ${\cal M}$
endowed with a metric $g_{\mu\nu}$.\\
In the Landau gauge and in the context of such a geometrical background
the gauge fixed action is:
\begin{equation}
\label{gaction}
\Sigma_{gf} = s \int_{\cal{M}} d^{4}x
\sqrt{ g} \left\{ g^{\mu\alpha} g^{\nu\beta}
\chi^{a}_{\alpha\beta} F^{a+}_{\mu\nu} -
g^{\mu\nu} ( \partial_{\mu} \bar{\varphi}^{a} ) \psi^{a}_{\nu} -
g^{\mu\nu} ( \partial_{\mu} \bar{c}^{a} ) A^{a}_{\nu} \right\},
\end{equation}
where s, is the nilpotent BRS operator, g is the determinant
of the metric $g_{\mu\nu}$ and $g^{\mu\nu}$ its inverse.
The other fields have their values in the adjoint representation
of the gauge group G, which is a Lie group, supposed
to be compact. $f^{abc}$ denots the structure constants.\\
We have also:
\begin{equation}
\label{dual}
F^{a+}_{\mu\nu} = \tfrac{1}{2} ( F^{a}_{\mu\nu} +
\tilde{F}^{a}_{\mu\nu} ),
\end{equation}
with
\begin{equation}
\label{sdual}
\tilde{F}^{a}_{\mu\nu} = \tfrac{1}{2} \epsilon_{\mu\nu\alpha\beta}
g^{\alpha\rho} g^{\beta\sigma} F^{a}_{\rho\sigma},
\end{equation}
and
\begin{equation}
\label{st.tensor}
F^{a}_{\mu\nu} = \partial_{\mu} A^{a}_{\nu} -
                 \partial_{\nu} A^{a}_{\mu} +
                 f^{abc} A^{b}_{\mu} A^{c}_{\nu},
\end{equation}
being the curvature associated with the gauge connection $A^{a}_{\mu}$.
The topological ghost is represented by the field $\psi^{a}_{\mu}$.\\
The full set of BRS transformations of the fields is:
\begin{equation}
\label{brs}
\begin{array}{ll}
s A^{a}_{\mu} = -(D_{\mu}c)^{a} + \psi^{a}_{\mu},              & \\
%                                                              & \\
s \psi^{a}_{\mu} = f^{abc} c^{b} \psi^{c}_{\mu} +
                   (D_{\mu}\varphi)^{a},                       & \\
%                                                              & \\
s c^{a} = \tfrac{1}{2} f^{abc} c^{b} c^{c} + \varphi^{a},      & \\
%                                                              & \\
s \varphi^{a} = f^{abc} c^{b} \varphi^{c},                     & \\
%                                                              & \\
s \bar{c}^{a} = b^{a},    &  s b^{a} = 0,                        \\
%                                                              & \\
s \bar{\varphi}^{a} = \eta^{a},   &  s \eta^{a} = 0,             \\
%                                                              & \\
s \chi^{a}_{\mu\nu} = B^{a}_{\mu\nu},   &  s B^{a}_{\mu\nu} = 0.    \\
%                                                               & \\

\end{array}
\end{equation}
$c^{a}$ and $\varphi^{a}$ are the two scalar ghost fields corresponding
respectively to $A^{a}_{\mu}$ and $\psi^{a}_{\mu}$.
We have also the three couples $(\chi^{a}_{\mu\nu},B^{a}_{\mu\nu})$,
$(\bar{\varphi}^{a},\eta^{a})$ and $(\bar{c}^{a},b^{a})$
each consisting of an antighost field and the corresponding
Lagrange multiplier.
The Lagrange multiplier $B^{a}_{\mu\nu}$ enforces the instantonic condition
\begin{equation}
F^{a+}_{\mu\nu} = 0,
\end{equation}
whereas $\eta^{a}$ and $b^{a}$ enforce the gauge fixings for the two fields
$\psi^{a}_{\mu}$ and $A^{a}_{\mu}$ respectively.
The next step is to introduce the external sources \cite{bec}
coupled to the BRS transformations which are non linear in the fields.
This external sources generate the external part of the action,
which we denote by $\Sigma_{ext}$:
\begin{equation}
\label{ext}
\Sigma_{ext} = \int_{\cal{M}} d^{4}x \left\{
\Omega^{a\mu} (D_{\mu}c)^{a} + \tau^{a\mu} (s \psi^{a}_{\mu}) +
\tfrac{1}{2} L^{a} f^{abc} c^{b} c^{c} +
D^{a} (s \varphi^{a}) \right\},
\end{equation}
where $\Omega^{a\mu}$ and $\tau^{a\mu}$ are contravariant vector
densities of weight one, whereas $L^{a}$ and $D^{a}$ are scalar
densities, also of weight one.
If we let the sources transform under BRS \cite{why} according to:
\begin{equation}
\label{brs.sources}
\begin{array}{ll}
s \tau^{a\mu} = \Omega^{a\mu},  ~~~~~~~&~~~~~~~ s \Omega^{a\mu} = 0,    \\
%                                                                     & \\
s D^{a} = L^{a},   ~~~~~~~&~~~~~~~  s L^{a} = 0.
\end{array}
\end{equation}
then $\Sigma_{ext}$ can be written in the simpler form:
\begin{equation}
\label{s.ext}
\Sigma_{ext} = s \int_{\cal{M}} d^{4}x \left(
\tau^{a\mu} (D_{\mu}c)^{a} +
\tfrac{1}{2} D^{a} f^{abc} c^{b} c^{c} \right),
\end{equation}
This leads to the total action:
\begin{equation}
\label{tot.action}
\Sigma = \Sigma_{gf} + \Sigma_{ext}
\end{equation}
which is just an s-exact expression:
\begin{eqnarray}
\label{s.action}
\Sigma & = & s \int_{\cal{M}} d^{4}x \left\{
             \sqrt{ g} \left( g^{\mu\alpha} g^{\nu\beta}
             \chi^{a}_{\alpha\beta} F^{a+}_{\mu\nu} -
         g^{\mu\nu} [ \partial_{\mu} \bar{\varphi}^{a} ] \psi^{a}_{\nu} -
         g^{\mu\nu} [ \partial_{\mu} \bar{c}^{a} ] A^{a}_{\nu} \right) +
             \right.\nonumber \\
       &+&   \left.\tau^{a\mu} (D_{\mu}c)^{a} +
             \tfrac{1}{2} D^{a} f^{abc} c^{b} c^{c} \right\},
\end{eqnarray}
Note that the operator s is nilpotent.\\
One can notice that in (\ref{s.action}) the metric $g_{\mu\nu}$
is introduced through a trivial BRS variation term, then the physical
objects are metric independent. In this case the metric is just a
gauge parameter.\\
The above arguments allow us to extend the BRS transformation \cite{ps}
by letting s acting on the metric \cite{lp} in the following way:
\begin{equation}
\label{s.metric}
\begin{array}{ll}
s g_{\mu\nu} = \hat{g}_{\mu\nu}, ~~~~~~~~~&~~~~~~~  s \hat{g}_{\mu\nu} = 0.
\end{array}
\end{equation}
Let us now, give in table (1) the dimensions an the ghost numbers
of the set of the fields described above:
\begin{table}[h]
\begin{center}
\begin{tabular}{|c|c|c|c|c|c|c|c|c|c|c|c|c|c|c|c|c|}\hline
      &$A^{a}_{\mu}$ &$\psi^{a}_{\mu}$ &$c^{a}$ &$\varphi^{a}$
      &$\chi^{a}_{\mu\nu}$ &$B^{a}_{\mu\nu}$ &$\bar{\varphi}^{a}$
      &$\eta^{a}$ &$\bar{c}^{a}$ &$b^{a}$ &$\Omega^{a\mu}$
      &$\tau^{a\mu}$ &$L^{a}$ &$D^{a}$ &$g_{\mu\nu}$
      &$\hat{g}_{\mu\nu}$ \\ \hline
dim   &1  &1  &0  &0  &2  &2  &2  &2  &2  &2  &3  &3
      &4  &4  &0  &0 \\ \hline
$\Phi\Pi$  &0  &1  &1  &2  &-1  &0  &-2  &-1  &-1  &0
           &-1  &-2  &-2  &-3  &0  &1 \\ \hline
\end{tabular}
\caption{Dimensions and ghost numbers of the fields}
\end{center}
\end{table}
\par
The next step is to generalise the vectors supersymmetry found
in \cite{why}, to a curved space-time.
To this end we propose the following set of transformations with respect
to a contravariant vector field $\xi^{\mu}$ of ghost number $+2$.
\begin{equation}
\label{susy}
\begin{array}{ll}
   \delta^{S}_{(\xi)} A^{a}_{\mu} = 0 ,                  ~~~~~~~&~~~~~~~
   \delta^{S}_{(\xi)} \psi^{a}_{\mu} = {\cal L}_{\xi} A^{a}_{\mu}, \\
%                                             &       \\
   \delta^{S}_{(\xi)} c^{a} = 0 ,                          ~~~~~~~&~~~~~~~
   \delta^{S}_{(\xi)} \varphi^{a} = {\cal L}_{\xi} c^{a} ,  \\
 %                                            &        \\
   \delta^{S}_{(\xi)} \chi^{a}_{\mu\nu} = 0 ,              ~~~~~~~&~~~~~~~
   \delta^{S}_{(\xi)} B^{a}_{\mu\nu} = {\cal L}_{\xi} \chi^{a}_{\mu\nu} , \\
%                                             &        \\
   \delta^{S}_{(\xi)} \bar{\varphi}^{a} = 0 ,              ~~~~~~~&~~~~~~~
   \delta^{S}_{(\xi)} \eta^{a} = {\cal L}_{\xi} \bar{\varphi}^{a} ,  \\
%                                             &        \\
   \delta^{S}_{(\xi)} \bar{c}^{a} =
                {\cal L}_{\xi} \bar{\varphi}^{a} ,       ~~~~~~~&~~~~~~~
   \delta^{S}_{(\xi)} b^{a} = {\cal L}_{\xi} \bar{c}^{a} -
                              {\cal L}_{\xi} \eta^{a} ,     \\
%                                             &         \\
  \delta^{S}_{(\xi)} \tau^{a\mu} = 0 ,                     ~~~~~~~&~~~~~~~
  \delta^{S}_{(\xi)} \Omega^{a\mu} =
             {\cal L}_{\xi} \tau^{a\mu} ,  \\
%                                              &          \\
  \delta^{S}_{(\xi)} D^{a} = 0 ,                           ~~~~~~~&~~~~~~~
  \delta^{S}_{(\xi)} L^{a} = {\cal L}_{\xi} D^{a} ,  \\
%                                               &          \\
  \delta^{S}_{(\xi)} g_{\mu\nu} = 0 ,                      ~~~~~~~&~~~~~~~
  \delta^{S}_{(\xi)} \hat{g}_{\mu\nu} = {\cal L}_{\xi} g_{\mu\nu} .
\end{array}
\end{equation}
It is straightforward to check that the anticommutator between the
BRS operator s and the superdiffeomorphisms operator $ \delta^{S}_{(\xi)} $
closes on the Lie derivatives in the direction of $ \xi^{\mu} $:
\begin{equation}
\label{onshell}
  \{ s , \delta^{S}_{(\xi)} \} = {\cal L}_{\xi}.
\end{equation}
At the functional level, the superdiffeomorphism transformations,
are implemented by means of the following
Ward operator $ {\cal W^{S}_{(\xi)}} $:
\begin{equation}
\label{superdiff.op}
{\cal W^{S}_{(\xi)}} = \int_{\cal M} d^{4}x \sum_{f}
                       \delta^{S}_{(\xi)} f
                       \frac{\delta}{\delta f},
\end{equation}
where f describes the set of all the fields transforming under
$ \delta^{S}_{(\xi)} $.\\
Now, we let the operator $ {\cal W^{S}_{(\xi)}} $ acting on the
total action (\ref{s.action}), and we get:
\begin{equation}
\label{h.b}
{\cal W^{S}_{(\xi)}} \Sigma = s \int_{\cal M} d^{4}x \lbrace
                              \sqrt{g} g^{\mu\nu} {\cal L}_{\xi} \lbrack
                              ( \partial_{\mu} \bar{\varphi}^{a} )
                              A^{a}_{\nu} \rbrack \rbrace .
\end{equation}
At this point, one can remark that in the special case where $\xi^{\mu}$
is a Killing vector $({\cal L}_{\xi} g_{\mu\nu} = 0)$ the right hand side
of (\ref{h.b}) will vanish. But in a general situation such breacking exist,
which is quadratic in the quantum fields. this will generate problems
at the quantum level, if one would try to quantise this model.
Fortunately, one can control such a non-linear breacking
by absorbing it in the original action by introducing two auxilliary fields
\cite{sym}, which we will call $ L^{\mu\nu} $ and $ M^{\mu\nu} $.
they are both symmetric contravariant tensors of rank two and
weight $ +1 $.\\
By adding to the original action (\ref{s.action}) the new term
\begin{equation}
\label{lm.action}
\Sigma_{L,M} = - \int_{\cal M} d^{4}x \lbrack
               L^{\mu\nu} \Xi_{\mu\nu} -
               M^{\mu\nu} s \Xi_{\mu\nu} \rbrack .
\end{equation}
with:
\begin{equation}
\label{xi}
\Xi_{\mu\nu} = ( \partial_{\mu} \bar{\varphi}^{a} )
               A^{a}_{\nu}
\end{equation}
we will get a vanishing right hand side in (\ref{h.b}).
So, our action takes the following form:
\begin{equation}
\label{new.action}
\Sigma = \Sigma_{gf} + \Sigma_{ext} + \Sigma_{L,M}.
\end{equation}
And in order to preserve the BRS invariance of this new action, we impose
the following transformations on the two fields $ L^{\mu\nu} $ and
$ M^{\mu\nu} $:
\begin{equation}
\begin{array}{ll}
s M^{\mu\nu} = L^{\mu\nu} , ~~~~~~~ & ~~~~~~~~   s L^{\mu\nu} = 0.
\end{array}
\end{equation}
Under the superdiffeomorphisms the auxilliary fields transform as:
\begin{equation}
\begin{array}{ll}
\delta^{S}_{(\xi)} M^{\mu\nu} = {\cal L}_{\xi} ( \sqrt{g} g^{\mu\nu} ), & \\
\delta^{S}_{(\xi)} L^{\mu\nu} = {\cal L}_{\xi} \lbrack
                                M^{\mu\nu} -  s ( \sqrt{g} g^{\mu\nu} )
                                \rbrack.
\end{array}
\end{equation}
guaranteeing both, the absence of the hard breacking in (\ref{h.b}) and
the validity of (\ref{onshell}).\\
So, now we have the following identity:
\begin{equation}
\label{superdiff}
{\cal W^{S}_{(\xi)}} \Sigma = 0
\end{equation}
where ${\cal W^{S}_{(\xi)}}$ is the operator defined above in
(\ref{superdiff.op}), and f includes now $L^{\mu\nu}$ and $M^{\mu\nu}$.
\newpage
In table $(2)$ we give the dimensions and the ghost numbers of the
fields $L^{\mu\nu}$ and $M^{\mu\nu}$ :
%\newpage
\begin{table}[h]
\begin{center}
\begin{tabular}{|c|c|c|}\hline
   &$L^{\mu\nu}$       &$M^{\mu\nu}$          \\    \hline
 dim                &0           &0            \\ \hline
$\Phi\Pi$           &2           &1             \\ \hline
\end{tabular}
\caption{dimensions and ghost numbers of the auxilliary fields.}
\end{center}
\end{table}
\par
At the functional level, the BRS invariance of the action is characterised
by the Slavnov identity:
\begin{eqnarray}
\label{sla.tay}
{\cal S}(\Sigma)&=& \int_{\cal M} d^{4}x \left(
                  \psi^{a}_{\mu}
                  \frac{\delta \Sigma}{A^{a}_{\mu}} -
                  \frac{\delta \Sigma}{\delta \Omega^{a\mu}}
                  \frac{\delta \Sigma}{\delta A^{a}_{\mu}} +
                  \varphi^{a}
                  \frac{\delta \Sigma}{\delta c^{a}} +
                  \frac{\delta \Sigma}{\delta L^{a}}
                  \frac{\delta \Sigma}{\delta c^{a}} +
                  \frac{\delta \Sigma}{\delta \tau^{a\mu}}
                  \frac{\delta \Sigma}{\delta \psi^{a}_{\mu}} +
                  \right.\nonumber \\
              &+& \left.\frac{\delta \Sigma}{\delta D^{a}}
                  \frac{\delta \Sigma}{\delta \varphi^{a}} +
                  \tfrac{1}{2} B^{a}_{\mu\nu}
                  \frac{\delta \Sigma}{\delta \chi^{a}_{\mu\nu}} +
                  \eta^{a}
                  \frac{\delta \Sigma}{\delta \bar{\varphi}^{a}} +
                  b^{a} \frac{\delta \Sigma}{\delta \bar{c}^{a}} +
                  \Omega^{a\mu}
                  \frac{\delta \Sigma}{\delta \tau^{a\mu}} +
                  \right.\nonumber \\
             &+&  \left.L^{a}
                  \frac{\delta \Sigma}{\delta D^{a}} +
                  \tfrac{1}{2} L^{\mu\nu}
                  \frac{\delta \Sigma}{\delta M^{\mu\nu}} +
                  \tfrac{1}{2} \hat{g}_{\mu\nu}
                  \frac{\delta \Sigma}{\delta g_{\mu\nu}} \right) = 0
\end{eqnarray}
from which one can write down the correponding linearised Slavnov operator:
%\newpage
\begin{eqnarray}
\label{sla.op}
{\cal S}_{\Sigma}&=& \int_{\cal M} d^{4}x \left(
                   \psi^{a}_{\mu}
                   \frac{\delta}{\delta A^{a}_{\mu}} -
                   \frac{\delta \Sigma}{\delta \Omega^{a\mu}}
                   \frac{\delta}{\delta A^{a}_{\mu}} -
                   \frac{\delta \Sigma}{\delta A^{a}_{\mu}}
                   \frac{\delta}{\delta \Omega^{a\mu}} +
                   \varphi^{a}
                   \frac{\delta}{\delta c^{a}} +
                   \frac{\delta \Sigma}{\delta L^{a}}
                   \frac{\delta}{\delta c^{a}} +
                   \frac{\delta \Sigma}{\delta c^{a}}
                   \frac{\delta}{\delta L^{a}} +
                   \right.\nonumber \\
               &+& \left.\frac{\delta \Sigma}{\delta \tau^{a\mu}}
                   \frac{\delta}{\delta \psi^{a}_{\mu}} +
                   \frac{\delta \Sigma}{\delta \psi^{a}_{\mu}}
                   \frac{\delta}{\delta \tau^{a\mu}} +
                   \frac{\delta \Sigma}{\delta D^{a}}
                   \frac{\delta}{\delta \varphi^{a}} +
                   \frac{\delta \Sigma}{\delta \varphi^{a}}
                   \frac{\delta}{\delta D^{a}} +
                   \tfrac{1}{2} B^{a}_{\mu\nu}
                   \frac{\delta}{\delta \chi^{a}_{\mu\nu}} +
                   \eta^{a}
                   \frac{\delta}{\delta \bar{\varphi}^{a}} +
                   \right.\nonumber \\
               &+& \left.b^{a}
                   \frac{\delta}{\delta \bar{c}^{a}} +
                   \Omega^{a\mu}
                   \frac{\delta}{\delta \tau^{a\mu}} +
                   L^{a}
                   \frac{\delta}{\delta D^{a}} +
                   \tfrac{1}{2} \hat{g}_{\mu\nu}
                   \frac{\delta}{\delta g_{\mu\nu}} +
                   \tfrac{1}{2} L^{\mu\nu}
                   \frac{\delta}{\delta M^{\mu\nu}} \right) .
\end{eqnarray}
Beside the BRS and the superdiffeomorphism invariance, the action
(\ref{new.action}) is also invariant under diffeomorphism transformations,
such that:
\begin{equation}
\label{diff}
{\cal W}^{D}_{(\varepsilon)} \Sigma = \int_{\cal M} d^{4}x
                                      \sum_{f} {\cal L}_{\varepsilon} f
                                      \frac{\delta \Sigma}{\delta f} = 0 .
\end{equation}
where the Ward operator for the diffeomorphisms is:
\begin{equation}
\label{diff.op}
{\cal W}^{D}_{(\varepsilon)} = \int_{\cal M} d^{4}x
                              \sum_{f} {\cal L}_{\varepsilon} f
                              \frac{\delta}{\delta f} .
\end{equation}
and f stands for the same fields as in (\ref{superdiff}),
$ \varepsilon^{\mu} $ is a
contravariant vector, the parameter of the diffeomorphisms transformations.\\
For reasons, which will be clear later we change the ghost number
of $ \varepsilon^{\mu} $ in such a way to render the
operator (\ref{diff.op}) fermionic.\\
The dimensions and the ghost numbers of the parameters $ \varepsilon^{\mu} $
and $ \xi^{\mu} $ are given in table $ (3) $: \\
\begin{table}[h]
\begin{center}
\begin{tabular}{|c|c|c|}\hline
    &$\varepsilon^{\mu}$   &$\xi^{\mu}$ \\ \hline
dim         &-1                   &-1     \\ \hline
$\Phi\Pi$   &1                    &2     \\ \hline
\end{tabular}
\caption{dimensions and ghost numbers of the parameters.}
\end{center}
\end{table}
\par
The action (\ref{new.action}) is invariant with respect to three symmetries,
the BRS, the diffeomorphisms and superdiffeomorphisms,
each symmetry generates a Ward operator and the three Ward operators
together display the following linear algebra:
\begin{equation}
\label{algebra}
\begin{array}{rcl}
\left\{ {\cal S}_{\Sigma},{\cal S}_{\Sigma} \right\} & = & 0,  \\
\left\{ {\cal S}_{\Sigma},{\cal W}^{D}_{(\varepsilon)} \right\} & = & 0, \\
\left\{ {\cal W}^{D}_{(\varepsilon)},{\cal W}^{D}_{(\varepsilon^{\prime})}
\right\} & = & -{\cal W}^{D}_{([\varepsilon,\varepsilon^{\prime}])}, \\
\left\{ {\cal S}_{\Sigma},{\cal W}^{S}_{(\xi)} \right\}
& = & {\cal W}^{D}_{(\xi)}, \\
\left\{ {\cal W}^{S}_{(\xi)}, {\cal W}^{D}_{(\varepsilon)} \right\}
& = & {\cal W}^{S}_{([\xi,\varepsilon])}, \\
\left\{ {\cal W}^{S}_{(\xi)},{\cal W}^{S}_{(\xi^{\prime})} \right\}
& = & 0.
\end{array}
\end{equation}
where we have denoted by $ [\ ,\ ] $ the graded Lie bracket, such that:
\begin{equation}
[\varepsilon,\varepsilon^{\prime}]^{\mu} =
 \varepsilon^{\lambda} \partial_{\lambda} \varepsilon^{\prime\mu} +
 \varepsilon^{\prime\lambda} \partial_{\lambda} \varepsilon^{\mu}.
\end{equation}
and
\begin{equation}
[\xi,\varepsilon]^{\mu} =
\xi^{\lambda} \partial_{\lambda} \varepsilon^{\mu} -
\varepsilon^{\lambda} \partial_{\lambda} \xi^{\mu}.
\end{equation}
Furthermore, the total action (\ref{new.action}) also satisfies: \\
(i) two gauge conditions:
\begin{equation}
\label{ga.con.1}
\frac{\delta \Sigma}{\delta b^{a}} = \partial_{\mu} ( \sqrt{g}
                                     g^{\mu\nu} A^{a}_{\nu} ) , \\
\end{equation}
\begin{equation}
\label{ga.con.2}
\frac{\delta \Sigma}{\delta \eta^{a}} = \partial_{\mu} ( \sqrt{g}
                                        g^{\mu\nu} \psi^{a}_{\nu} ) +
                           \partial_{\mu} ( M^{\mu\nu} A^{a}_{\nu} ) .
\end{equation}
(ii) two antighost equations:
\begin{equation}
\label{an.gho.1}
\frac{\delta \Sigma}{\delta \bar{c}^{a}} - \partial_{\mu} ( \sqrt{g}
g^{\mu\nu} \frac{\delta \Sigma}{\delta \Omega^{a\nu}} ) = -
\partial_{\mu} [ s ( \sqrt{g} g^{\mu\nu} ) A^{a}_{\nu} ] - \partial_{\mu}
 [ \sqrt{g} g^{\mu\nu} \psi^{a}_{\nu} ] ,
\end{equation}
\begin{equation}
\label{an.gho.2}
\frac{\delta \Sigma}{\delta \bar{\varphi}^{a}} -
\partial_{\mu} [ \sqrt{g} g^{\mu\nu}
\frac{\delta \Sigma}{\delta \tau^{a\nu}} ] - \partial_{\mu} [
M^{\mu\nu} \frac{\delta \Sigma}{\delta \Omega^{a\nu}} ] =
\partial_{\mu} [ s ( \sqrt{g} g^{\mu\nu} ) \psi^{a}_{\nu} ] +
\partial_{\mu} ( L^{\mu\nu} A^{a}_{\nu} ) -
\partial_{\mu} ( M^{\mu\nu} \psi^{a}_{\nu} ) .
\end{equation}
(iii) two ghost equations, present in the Landau type gauge \cite{bps}:
\begin{equation}
\label{gho.1}
\int_{\cal M} d^{4}x ( \frac{\delta \Sigma}{\delta \varphi^{a}} -
f^{abc} \bar{\varphi}^{b} \frac{\delta \Sigma}{\delta b^{c}} ) =
\int_{\cal M} d^{4}x f^{abc} ( \tau^{b\mu} A^{c}_{\mu} +
D^{b} c^{c} ) ,
\end{equation}
and
\begin{equation}
\label{gho.2}
{\cal G}^{a} \Sigma = \Delta^{a}_{\cal G}
\end{equation}
where ${\cal G}^{a}$ writes as:
\begin{equation}
\label{gho.op2}
{\cal G}^{a} = \int_{\cal M} d^{4}x \left(
               \frac{\delta}{\delta c^{a}} +
               \tfrac{1}{2} f^{abc} \chi^{b}_{\mu\nu}
               \frac{\delta}{\delta B^{c}_{\mu\nu}} +
               f^{abc} \bar{\varphi}^{b}
               \frac{\delta}{\delta \eta^{c}} +
               f^{abc} \bar{c}^{b}
               \frac{\delta}{\delta b^{c}} \right)
\end{equation}
and $\Delta^{a}_{\cal G}$ has the following form:
\begin{equation}
\Delta^{a}_{\cal G} = \int_{\cal M} d^{4}x f^{abc} \left(
                      D^{b} \varphi^{c} -
                      \Omega^{b\mu} A^{c}_{\mu} -
                      \tau^{b\mu} \psi^{c}_{\mu} -
                      L^{b} c^{c} \right)
\end{equation}
By commuting the ghost equation (\ref{gho.1}) with the Slavnov
identity (\ref{sla.tay}) we get a further constraint \cite{why}:
\begin{equation}
\label{gho.3}
{\cal F}^{a} \Sigma = \int_{\cal M} d^{4}x f^{abc} (
A^{b}_{\mu} \Omega^{c\mu} +
\tau^{c\mu} \psi^{b}_{\mu} +
L^{c} c^{b} + D^{b} \varphi^{c} ).
\end{equation}
where $ {\cal F}^{a} $ denote the following operator:
\begin{eqnarray}
\label{gho.op3}
{\cal F}^{a}&=& \int_{\cal M} d^{4}x \left( \frac{\delta}{\delta c^{a}} -
    f^{abc} \bar{\varphi}^{b} \frac{\delta}{\delta \bar{c}^{c}} -
    f^{abc} A^{b}_{\mu} \frac{\delta}{\delta \psi^{c}_{\mu}} -
    \right.\nonumber \\
&-& \left. f^{abc} \tau^{b\mu} \frac{\delta}{\delta \Omega^{c\mu}} -
    f^{abc} c^{b} \frac{\delta}{\delta \varphi^{c}} -
    f^{abc} D^{b} \frac{\delta}{\delta L^{c}} +
    f^{abc} \eta^{b} \frac{\delta}{\delta b^{c}} \right).
\end{eqnarray}
Further, by anticommuting (\ref{gho.2}) with (\ref{sla.tay}) we get:
\begin{equation}
\label{gho.4}
{\cal R}^{a} \Sigma = 0
\end{equation}
with:
\begin{equation}
\label{gho.op4}
{\cal R}^{a} = \int_{\cal M} d^{4}x
               \sum_{\omega} f^{abc} \omega^{b}
               \frac{\delta}{\delta \omega^{c}}.
\end{equation}
Where $\omega$ stands for all fields possessing a gauge index.\\
We conclude this section by noting that the above linear algebra
(\ref{algebra})
generated by the Slavnov operator and the Ward operators
for diffeomorphisms and superdiffeomorphisms, has the same structure
as in the case of the Chern-Simons theory \cite{lp} and the three
dimensional BF model \cite{mwosz} on a curved manifold.

\section{Quantization}

So far we have considered only the classical analysis of the topological
Yang-Mills on an arbitrary curved four dimensional space-time.
Now, we would like to discusse the possibility to describe the theory
at the quantum level.\\
The first step is the analysis of the stability and the search of possible
counterterms. The most general counterterm $ \Delta $ is an integrated
local field polynomial of vanishing dimension and ghost number, such that
the perturbed action takes the form:
\begin{equation}
{\Sigma}^{\prime} = \Sigma + \Delta.
\end{equation}
It is easy to check that the perturbation $ \Delta $ obeys the following
constraints:\\
%\newpage
\begin{equation}
\label{c1}
\frac{\delta \Delta}{\delta b^{a}} = 0,
\end{equation}
\begin{equation}
\label{c2}
\frac{\delta \Delta}{\delta \eta^{a}} = 0,
\end{equation}
\begin{equation}
\label{c3}
\frac{\delta \Delta}{\delta \bar{c}^{a}} - \partial_{\mu} (
\sqrt{g} g^{\mu\nu} \frac{\delta \Delta}{\delta \Omega^{a\nu}} )
= 0,
\end{equation}
\begin{equation}
\label{c4}
\frac{\delta \Delta}{\delta \bar{\varphi}^{a}} - \partial_{\mu} (
\sqrt{g} g^{\mu\nu} \frac{\delta \Delta}{\delta \tau^{a\nu}} ) -
\partial_{\mu} ( M^{\mu\nu} \frac{\delta \Delta}{\delta \Omega^{a\nu}} )
= 0,
\end{equation}
\begin{equation}
\label{c5}
\int_{\cal M} d^{4}x
\frac{\delta \Delta}{\delta \varphi^{a}} = 0,
\end{equation}
\begin{equation}
\label{c6}
{\cal G}^{a} \Delta = 0,
\end{equation}
\begin{equation}
\label{c7}
{\cal F}^{a} \Delta = 0,
\end{equation}
\begin{equation}
\label{c8}
{\cal R}^{a} \Delta = 0,
\end{equation}
\begin{equation}
\label{c9}
{\cal S}_{\Sigma} \Delta = 0,
\end{equation}
\begin{equation}
\label{c10}
{\cal W}^{D}_{(\varepsilon)} \Delta = 0,
\end{equation}
\begin{equation}
\label{c11}
{\cal W}^{S}_{(\xi)} \Delta = 0.
\end{equation}
The first two equations (\ref{c1}) and (\ref{c2}) imply that $\Delta$
is independent of the fields $b^{a}$ and $\eta^{a}$. From the next two
equations (\ref{c3}) and (\ref{c4}), it follows that $\Delta$ depends
on $\bar{c}^{a}$,
$\Omega^{a\mu}$, $\bar{\varphi}^{a}$ and $\tau^{a\mu}$ only through the
two combinations:
\begin{equation}
\tilde{\Omega}^{a\mu} = \Omega^{a\mu} - \sqrt{g} g^{\mu\nu}
                        \partial_{\nu} \bar{c}^{a} -
                        M^{\mu\nu} \partial_{\nu} \bar{\varphi}^{a}
\end{equation}
and,
\begin{equation}
\tilde{\tau}^{a\mu} = \tau^{a\mu} - \sqrt{g} g^{\mu\nu}
                      \partial_{\nu} \bar{\varphi}^{a}.
\end{equation}
The last tree equations (\ref{c9}), (\ref{c10}) and (\ref{c11}) can be put
together \cite{bbbcd} in such a way
that they generate a single cohomology problem \cite{lp} and \cite{mwosz}:
\begin{equation}
\label{co.pro}
\delta \Delta = 0
\end{equation}
where $ \delta $ denots a nilpotent operator:
\begin{equation}
\label{nil}
\delta^{2} = 0
\end{equation}
given by:
\begin{equation}
\label{zzz}
\delta = {\cal S}_{\Sigma} +
         {\cal W}^{D}_{(\varepsilon)} +
         {\cal W}^{S}_{\xi} +
         {\cal U}_{(\xi)} +
         {\cal V}_{(\varepsilon)}.
\end{equation}
with:
\begin{equation}
{\cal U}_{(\xi)} = \int_{\cal M} d^{4}x [ \varepsilon, \xi ]^{\mu}
                   \frac{\delta}{\delta \xi^{\mu}},
\end{equation}
and
\begin{equation}
{\cal V}_{(\varepsilon)} = \int_{\cal M} d^{4}x \lbrace \tfrac{1}{2}
                           [ \varepsilon , \varepsilon ]^{\mu} -
                           \xi^{\mu} \rbrace
                           \frac{\delta}{\delta \varepsilon^{\mu}}.
 \end{equation}
In order to solve the cohomology problem (\ref{co.pro}) it is useful
to introduce a filtring operator
\footnote{All fields have homogeneity degree one,
except $c^{a}$,
$\varphi^{a}$, $\tau^{a\mu}$ and $\Omega^{a\mu}$ to which we attribute
homogeneity degree two.}
\cite{dix}, which will induce  a splitting of the operator $ \delta $ as:
\begin{equation}
\label{split}
\delta = \delta_{0} + \delta_{1} + ... + \delta_{N}
\end{equation}
where the $\delta_{n}$ increase the homogeneity degree by n unites.
This will reduce the original cohomology problem (\ref{co.pro}) to
a simpler one involving the operator $\delta_{0}$:
\begin{equation}
\label{co.pro1}
\delta_{0} \Delta = 0,
\end{equation}
the nilpotency of the whole operator $\delta$ (\ref{nil}) implies:
\begin{equation}
\delta_{0} \delta_{0} = 0
\end{equation}
%It is important to stress that the cohomology of $\delta$ is
%isomorphic to a subspace of the cohomology of $\delta_{0}$
%
%It is important to stress that the
%the set of solutions of (\ref{co.pro}) modulo trivial
%$\delta$--cocycles
%is isomorphic to a subspace of the set of solutions of $\ref{co.pro1}$
%modulo trivial $\delta_{0}$--cocycles.
Now, let us give the explicite expression of the nilpotent operator
$\delta_{0}$:
\begin{eqnarray}
\label{xxx}
\delta_{0} & = & \int_{\cal M} d^{4}x \left(
             \psi^{a}_{\mu}
             \frac{\delta}{\delta A^{a}_{\mu}} +
             \varphi^{a}
             \frac{\delta}{\delta c^{a}} +
             \tfrac{1}{2} B^{a}_{\mu\nu}
             \frac{\delta}{\delta \chi^{a}_{\mu\nu}} +
             \eta^{a}
             \frac{\delta}{\delta \bar{\varphi}^{a}} +
             b^{a}
             \frac{\delta}{\delta \bar{c}^{a}} +
             \right.\nonumber \\
      & + &  \left.\Omega^{a\mu}
             \frac{\delta}{\delta \tau^{a\mu}} +
             L^{a}
             \frac{\delta}{\delta D^{a}}     +
             \tfrac{1}{2} \hat{g}_{\mu\nu}
             \frac{\delta}{\delta g_{\mu\nu}} +
             \tfrac{1}{2} L^{\mu\nu}
             \frac{\delta}{\delta M^{\mu\nu}} -
             \xi^{\mu}
             \frac{\delta}{\delta \varepsilon^{\mu}}
             \right)
 \end{eqnarray}
which implies the triviality of the cohomology of
$\delta_{0}$: all the fields transform as a doublet under $\delta_{0}$.
This means that the cohomology of $\delta_{0}$, which is the space of
all solutions of (\ref{co.pro1}) modulo cocycles $\delta_{0} \bar{\Delta}$,
is empty.
%Now it is important to stress that the
%the set of solutions of (\ref{co.pro}) modulo trivial
%$\delta$--cocycles
%is isomorphic to a subspace of the set of solutions of $\ref{co.pro1}$
%modulo trivial $\delta_{0}$--cocycles.
Now, it is important to stress that the cohomology of $\delta$ is
isomorphic to a subspace of the cohomology of $\delta_{0}$
(see for instance \cite{dix}). This implies the triviality of
the cohomology of $\delta$. \\
%the only possible solution of $(\ref{co.pro})$
%is a trivial $\delta$--cocycle.\\
This last result leaves us with the most general counterterm, which is
a $\delta$-exact variation solving (\ref{co.pro}):
\begin{equation}
\label{sol}
\Delta = \delta \hat{\Delta}.
\end{equation}
Then the most general counterterm obeying the constraints
(\ref{c1}) -- (\ref{c8}), takes the  the following form:
%\newpage
\begin{eqnarray}
\label{coun.ter}
\Delta & = & \delta \int_{\cal M} d^{4}x \left(
           \alpha_{1} A^{a}_{\mu} [ \Omega^{a\mu} - \sqrt{g} g^{\mu\nu}
           \partial_{\nu} \bar{c}^{a} - M^{\mu\nu} \partial_{\nu}
           \bar{\varphi}^{a} ] +
           \alpha_{1} \psi^{a}_{\mu} [ \tau^{a\mu} - \sqrt{g} g^{\mu\nu}
           \partial_{\nu} \bar{\varphi}^{a} ] +
           \right.\nonumber \\
      &+&  \left.\alpha_{2} f^{abc} \sqrt{g} g^{\mu\rho} g^{\nu\sigma}
           A^{a}_{\mu} A^{b}_{\nu} \chi^{c}_{\rho\sigma} +
           \alpha_{3} \frac{1}{\sqrt{g}} g_{\rho\sigma} M^{\rho\sigma}
           A^{a}_{\mu} [ \tau^{a\mu} - \sqrt{g} g^{\mu\nu} \partial_{\nu}
           \bar{\varphi}^{a}] +
           \right.\nonumber \\
      &+&  \left.\alpha_{4} \frac{1}{\sqrt{g}} g_{\rho\sigma}
           M^{\mu\sigma} A^{a}_{\mu} [ \tau^{a\rho} - \sqrt{g}
           g^{\rho\nu} \partial_{\nu} \bar{\varphi}^{a} ] +
           \alpha_{5} g^{\mu\sigma} \hat{g}_{\mu\nu} A^{a}_{\sigma}
           [ \tau^{a\nu} - \sqrt{g} g^{\nu\rho} \partial_{\rho}
           \bar{\varphi}^{a} ] +
           \right.\nonumber \\
      &+&  \left.\alpha_{6} \sqrt{g} g^{\mu\rho} g^{\nu\sigma}
           A^{a}_{\rho} \partial_{\sigma} \chi^{a}_{\mu\nu} +
           \alpha_{7} [ \tau^{a\mu} - \sqrt{g} g^{\mu\nu} \partial_{\nu}
           \bar{\varphi}^{a} ] \partial_{\mu} c^{a} +
           \right.\nonumber \\
      &+&  \left.\alpha_{8} g_{\mu\nu} M^{\mu\nu} g^{\rho\lambda}
           g^{\sigma\kappa} \chi^{a}_{\lambda\kappa}
           \chi^{a}_{\rho\sigma} +
           \alpha_{9} M^{\mu\rho} g^{\nu\lambda}
           \chi^{a}_{\lambda\mu} \chi^{a}_{\nu\rho} +
           \right.\nonumber \\
      &+&  \left.\alpha_{10} \sqrt{g} g^{\mu\nu} \hat{g}_{\mu\nu}
           g^{\rho\lambda} g^{\sigma\kappa} \chi^{a}_{\lambda\kappa}
           \chi^{a}_{\rho\sigma} +
           \alpha_{11} \sqrt{g} g^{\mu\rho} \hat{g}_{\mu\sigma}
           g^{\nu\lambda} g^{\sigma\kappa} \chi^{a}_{\lambda\kappa}
           \chi^{a}_{\nu\rho} \right).
\end{eqnarray}
where the $\alpha_{i}$ are constant coefficients.
Let us now remark that the parameters of diffeomorphisms and
superdiffeomorphisms are present explicitly in the expression of the
$\delta$ operator (\ref{zzz}).
And as the action (\ref{new.action}) is independent of $\xi^{\mu}$ and
$\varepsilon^{\mu}$ the above counterterm has also to be independent
of such parameters. This further requirement will reduce (\ref{coun.ter})
to:
\begin{eqnarray}
\label{re.coun}
\Delta & = & {\cal S}_{\Sigma} \int_{\cal M} d^{4}x \left(
           \alpha_{1} A^{a}_{\mu} [ \Omega^{a\mu} - \sqrt{g} g^{\mu\nu}
           \partial_{\nu} \bar{c}^{a} - M^{\mu\nu} \partial_{\nu}
           \bar{\varphi}^{a} ] +
           \alpha_{1} \psi^{a}_{\mu} [ \tau^{a\mu} - \sqrt{g} g^{\mu\nu}
           \partial_{\nu} \bar{\varphi}^{a} ] +
           \right.\nonumber \\
      &+&  \left.\alpha_{2} f^{abc} \sqrt{g} g^{\mu\rho} g^{\nu\sigma}
           A^{a}_{\mu} A^{b}_{\nu} \chi^{c}_{\rho\sigma} +
           \alpha_{3} \sqrt{g} g^{\mu\rho} g^{\nu\sigma} A^{a}_{\rho}
           \partial_{\sigma} \chi^{a}_{\mu\nu} +
           \right.\nonumber \\
      &+&  \left.\alpha_{4} [ \tau^{a\mu} - \sqrt{g} g^{\mu\nu}
           \partial_{\nu} \bar{\varphi}^{a} ] \partial_{\mu} c^{a} +
           \alpha_{5} g_{\mu\nu} M^{\mu\nu} g^{\rho\lambda}
           g^{\sigma\kappa} \chi^{a}_{\lambda\kappa}
           \chi^{a}_{\rho\sigma} +
           \right.\nonumber \\
      &+&  \left.( - 4 \alpha_{5} ) M^{\mu\rho}
           g^{\nu\lambda} \chi^{a}_{\lambda\mu}
           \chi^{a}_{\nu\rho} +
           \alpha_{5} \sqrt{g} g^{\mu\nu} \hat{g}_{\mu\nu}
           g^{\rho\lambda} g^{\sigma\kappa} \chi^{a}_{\lambda\kappa}
           \chi^{a}_{\rho\sigma} +
           \right.\nonumber \\
      &+&  \left.( - 4 \alpha_{5} ) \sqrt{g} g^{\mu\rho}
           \hat{g}_{\mu\sigma} g^{\nu\lambda} g^{\sigma\kappa}
           \chi^{a}_{\lambda\kappa} \chi^{a}_{\nu\rho} \right)
\end{eqnarray}
where we have used the equality (\ref{zzz}). Now we can observe that
in the flat space limit (\ref{re.coun}) reduces to the counterterm
already found in \cite{why}. \\
The other important point to discuss in the framework of renormalization
theory, is the question of the existence of anomalies.\\
In our case, and due to the triviality of the cohomology of $\delta$,
the Slavnov identity as well as the two Ward identities of
diffeomorphisms and superdiffeomorphisms are not anomalous. Then they can
be promoted to the quantum level.\\
Concerning the constraints (\ref{c1}) -- (\ref{c4}) they
can be also promoted to the quantum level by using standard arguments
\cite{pr}.\\
The two ghost equations possess a linear breacking, and they are
also valid at the quantum level \cite{bps}.\\
This will conclude our proof of the ultraviolet renormalization of the
topological Yang-Mills considered in a curved space-time, topologically
equivalent to a flat space-time and
endowed with an asymptotically flat metric.

\section{Conclusion}

We could extend
the renormalization of the topological
Yang-Mills theory, already present and carried out in the flat space, to a
curved manifold by using the strategy of \cite{lp}, \cite{mwosz}.
Where the manifold has to be topologically equivalent to a flat
space and possessing an assymptotically flat metric in order to guarantee
the existence of a substraction scheme, which has not to be specified
in the framework of the
algebraic renormalization.
The most general counterterm was also constructed, and
the results of \cite{why} is in accordence with our analysis
in flat space--time limit.

\section*{Acknowledgements}

The author would like to thank M.W. de Oliveira for useful discussion,
O. Piguet for a critical reading of the manuscript, S. Emery and
M. Schweda for pertinent remarks. The author is also grateful
to the ``Fonds zur F\"orderung der
Wissenschaftlichen Forschung'' for financial support.
%\newpage

\end{document}